\documentclass[reprint,superscriptaddress,twocolumn,nofootinbib,notitlepage]{revtex4}
\usepackage[utf8x]{inputenc}
\usepackage{amsmath,amssymb}

\usepackage{graphicx}
\usepackage[mathscr]{eucal}
\usepackage{color}

\usepackage{graphicx}
\usepackage{bm}
\usepackage{dcolumn}
\usepackage[mathscr]{eucal}
\usepackage{color}

\newcommand{\bea}{\begin{eqnarray}}	
\newcommand{\eea}{\end{eqnarray}}
\newcommand{\be}{\begin{equation}}	
\newcommand{\ee}{\end{equation}}

\def\be{\begin{equation}} \def\ee{\end{equation}}

\DeclareMathOperator{\e}{\mathrm{e}}

\def\be{\begin{equation}}
\def\ee{\end{equation}}

\newcommand{\beqa}{\begin{eqnarray}}
\newcommand{\eeqa}{\end{eqnarray}}

\newcommand{\da}{\dagger}

 \newcommand{\ket}[1]{| {#1} \rangle}
 \newcommand{\inner}[2]{\langle {#1}| {#2} \rangle}

\begin{document}

\title{Geometric phases and cyclic isotropic cosmologies}
\author{ Leonardo Banchi}
\email{l.banchi@ucl.ac.uk}
\affiliation{Department of Physics, University College London, Gower Street, London WC1E 6BT, UK}
\author{Francesco Caravelli}
\email{f.caravelli@invenialabs.co.uk}
\affiliation{Invenia Labs, 27 Parkside Place, Parkside, Cambridge CB1 1HQ, UK}
\affiliation{London Institute of Mathematical Sciences, 35a South Street, London W1K 2XF, UK}

\begin{abstract}
In the present paper we study the evolution of the modes of a scalar field in a cyclic cosmology. In order to keep the discussion clear, we study the features of a scalar field in a toy model, a Friedman-Robertson-Walker universe with a periodic scale factor, in which the universe expands, contracts and bounces infinite times, in the approximation in which the dynamic features of this universe are driven by some external factor, without the backreaction of the scalar field under study. In particular, we show that particle production exhibits features of the cyclic cosmology.  Also, by studying the Berry phase of the scalar field, we show that contrarily to what is commonly believed, the scalar field carries information from one bounce to another in the form of a global phase which occurs to be generically non-zero. {The Berry phase is then evaluated numerically in the case of the effective Loop Quantum Cosmology closed Universe. We observe that Berry's phase is nonzero, but that in the quantum regime the particle content is non-negligible.}
\end{abstract}


\maketitle

\section{Introduction}
Our current understanding of the Universe and of its history relies on several puzzling observational features. 
In fact, soon after the discovery of the Cosmological Microwave Background (CMB), theoretical physicists realized that 
the early universe had to follow a rapid expanding phase. Inflation was the first candidate for generating such expansion, 
and a more refined version of inflation is still at the basis of the current standard cosmological model.
There are compelling evidences for a rapid expanding phase, which have culminated in the recent results of the BICEP and PLANCK collaborations
\cite{Planck1,Planck2,Planck3}. The current standard model of the universe involves an inflation field, but current experimental results, meanwhile ruling out many proposed models, seem not to exclude several other proposals alternative to inflation\cite{Lehners}.
Inflation has the advantage of the Occam's razor, as it accounts for all the puzzles involved with our current understanding of the Universe, such as its flatness and acceleration, its entropy and the horizon problem in one shot.

However, an alternative and interesting proposal relies on the idea of cyclicality of the Universe, a concept almost as old as General Relativity. Cosmological bounce models have a long history, with its first proposal put forward by Tolman \cite{tolman34}, and have been proposed in several different contexts \cite{osccuni}. In particular, it has been shown that
oscillating cosmological universes can provide a solution to both flatness and horizon problems, as long as maximum of the expansion in each cycle is increased in the next cycle.

Quantum gravitational phenomena play a role only in the extremely early Universe. In this scenario, quantum gravity is the force that makes the bounce occur. In fact, a Big Bounce has been observed in several quantum gravitational cosmological theories, as for instance Loop Quantum Cosmology \cite{lqc1,lqc2,lqc3}, Epkyrotic Strings\cite{ekp1,ekp2,ekp3}, Asymptotically safe non-local gravity \cite{Calcagni1} and in General Relativity with matter-torsion interaction \cite{bouncegr1} just to name few. 


In the present paper, we ask ourselves whether such bouncing phenomena has some physical, long-term effects at the quantum level. At the classical level, we find this unlikely to occur. However, one could argue that considering the Universe conformal factor as an external parameter, in a bouncing cosmology 
a Berry phase might emerge at the quantum level. This is the scenario we are
interested in: does a scalar field maintain a quantum memory from previous bounces? We answer this question in the modest nonphysical scenario of a scalar field in a bouncing cosmology,  in which the scalar field does not lead to any backreaction on the underlying metric; yet, this toy model features many of the stylized characteristics of bouncing cosmologies \cite{birreldavies,mukwin}. The Berry phase has been already considered in the cosmological setting, by for the case of inflationary models \cite{bpm}.

In general, the Berry phase is not a measurable quantity if the scalar field is
isolated. However, if the scalar field is entangled with other fields as
considered for instance in \cite{Albrecht}, then a Berry phase is measurable. To
see this, let us consider a scalar field $\phi$, in the state $|\phi,n\rangle$
at the beginning of the $n$-th cycle and let $\mathcal U_n$ be the propagator
which maps $|n\rangle\to|n{+}1\rangle$. If the scalar field's state does not change 
from one cycle to the other, i.e. $\mathcal U_n |\phi,n\rangle= |\phi,n{+}1\rangle$, 
then the bounce produces no observable effects. However, even though the field
is unchanged after one cycle, its state can pick a geometric phase $\mathcal
U_n|\phi,n\rangle=e^{i \alpha} |\phi,n{+}1\rangle$ which in principle can be
measured as in interferometry. For instance, for the sake of argument, 
we consider the simplest case of two scalar fields $\phi$ and $\xi$ which 
are in the entangled state 
$|\phi_{\rm
universe},n\rangle=\rho_1 |\phi,\xi,n\rangle +\rho_2 |\tilde \phi,\tilde
\xi,n\rangle$, although our argument can be extended to more complicated
cases. 
After a cycle, because of the Berry's phase, the state is a different one
$\rho_1 e^{i \alpha} |\phi,\xi,n{+}1\rangle +\rho_2 e^{i \tilde \alpha} |\tilde
\phi,\tilde \xi,n{+}1\rangle$. 
We argue that this phenomenon is poorly described by analysing, for instance,
the particle content coefficients only, in particular if these are calculated
perturbatively and can vanish identically, and that a nonperturbative approach
is more appropriate for calculating these phases. 
We will describe how to define a 
simple observable in Section \ref{berryphase}, and we will provide an
exact formula for the operator $\mathcal U_n$. {We will provide both explicit formulae for 
specific scale factor functions and provide general results for smooth scale factors which are periodic.}

The paper is organized as follows. In section \ref{scalarfield} we introduce the scalar field in curved spacetime, the settings and the notation, assuming that the Universe is cyclic. In section \ref{partprod} we study particle production in the toy model we introduce, both in a Kronig-Penney approximation and solving numerically the mode equation. In section \ref{berryphase}
we study the Berry Phase. Conclusions follow.


\section{Scalar field in a cyclic toy universe}
\label{scalarfield}
We are interested in the quantum features of a scalar field in 
a Friedman-Robertson-Walker (FRW) space time defined by the metric 
\begin{equation}
 ds^2=dt^2-a^2(t) (dx^2+dy^2+dz^2),
\end{equation}
which describes an evolving isotropic flatspace with a dynamic scale factor $a(t)$. 
In order to simplify many of the calculations, we introduce the conformal time $\eta=\int^t \frac{dt^\prime}{a(t^\prime)}$ coordinate. In these coordinates, the metric reads:
\begin{equation}
  ds^2=a(\eta)^2[d\eta^2- (dx^2+dy^2+dz^2)],
  \label{e.metriccoord}
\end{equation}
where $a(\eta)\equiv a(t(\eta))$ is the scale factor in the new coordinate.

As a simplifying assumption, we consider a scale factor $a(t)$ which oscillates according to a local time and that $a(t)>0$  as considered in many nonsingular bouncing
  cosmology models \cite{ekp1,lqc1,Calcagni1}, and that can be expanded in Fourier series.
 We assume that space is locally  flat, i.e. that the curvature of space is zero. In addition to this, we assume that the expansion and contracting phases of the universe last an equal amount of conformal time, which is a scenario ruled out from the physical viewpoint.

  In modelling the universe with these features, we were inspired by the two
  major current theories of bouncing cosmologies: the ekpyrotic scenario
  \cite{ekp1,ekp2,ekp3} and loop quantum cosmology \cite{lqc1,lqc2,lqc3, asht1, merc1}.
  Meanwhile the former has been affected with the recent results from the Planck
  collaboration \cite{Planck1}, the latter has shown to be a long-living toy
  model. In general, the
  ekpyrotic scenario relies on an asymmetric expanding and contracting phases,
  differently from our case, meanwhile the loop quantum cosmology approach is
  symmetric in its expanding and contracting phases.  It has been argued however that the ekpyrotic scenario suffers of singularities \cite{cail1}. Although violating the Tolman principle \cite{tolman34}, one could claim that a bounce occurring in a quantum gravity scenario might require an extension of the principles of thermodynamics, similarly to what happens for black holes. Thus, in this respect, we are more interested in understanding some of the features of a quantum field when the universe is cyclical.
 This simple approach has the advantage that one can carry on several
 calculations analytically. 

The action for a scalar field on a curved background is 
\begin{equation}
 S=\int dx^4 \sqrt{-g} [g^{\mu \nu}\partial_\mu \Phi \partial_\nu \Phi^*-V(g) \Phi \Phi^*]~,
 \label{e.action}
\end{equation}
where we consider the curvature dependent potential $V(g) = m^2+\xi R$, 
being $R=6\frac{a''(\eta)}{a(\eta)^3}$  for the metric \eqref{e.metriccoord} 
The equations of motion for a scalar field of this form are given by:
\begin{equation}
\frac{1}{\sqrt{-g}} \partial_\mu (g^{\mu \nu} \sqrt{-g} \partial_\nu \Phi)+V(g) \Phi=0 
\end{equation}
It is convenient to work with a field re-definition useful in the conformal time coordinates, 
$\Phi(x)=\frac{1}{\sqrt{(2\pi)^3a^2(\eta)}} e^{- i \vec k \cdot \vec x}
\chi(\eta)$, 
in which the equations of motion for the scalar field are given by:
$ [\frac{d^2}{d\eta^2}+\omega_k^2(\eta)]\chi(\eta)=0$,
with $\omega_k^2(\eta)=k^2+m^2 a^2(\eta)+(\xi-\frac{1}{6})R a^2(\eta)$.
If one works with conformally coupled fields, $\xi=1/6$, then the dependence of
the curvature on the modes disappears. 
  Moreover, we assume space isotropy, so all the quantities only depend on 
  $k=|\vec k|$.

  The classical field can be quantized by introducing bosonic creation and 
annihilation operators ($\hat{a}^\dagger_{\vec k}, \hat{a}_{\vec k}$) 
\cite{winitzky}: the expansion
\begin{align}
  \hat \Phi(\vec x, \eta) = \int &\frac{d^3 \vec k}{\sqrt{2(2\pi)^3 a(\eta)^2}}
  \times \nonumber \\&\times
  \left[
  \e^{i\vec k \cdot \vec x} \phi^*_k(\eta) \hat{a}_{\vec k} +
  \e^{ {-}i\vec k \cdot \vec x} \phi_k(\eta) \hat{a}^\da_{\vec k} 
\right]~,
  \label{Quantization}
\end{align}
defines the mode operators from the quantum field $ \hat \Phi(\vec x, \eta) $
and the mode functions satisfy the equation 
\begin{equation}
 \left[\frac{d^2}{d\eta^2}+\omega_k^2(\eta)\right]\phi_k(\eta)=0~.
 \label{EQDIF}
\end{equation}
Once the mode operators are determined, the vacuum state $\ket\Omega $ 
can be defined as the eigenstate of the 
annihilation operators with eigenvalue 0, i.e. $\hat{a}_{\vec k} \ket \Omega = 0$. 
However, different solutions of \eqref{EQDIF} define different mode operators
and thus different vacua. The physical vacuum 
$\ket{\Omega(\eta_0)}$ at a certain conformal time 
$\eta_0$ corresponds to the (instantaneous) lowest energy state and 
it is given by the solution of \eqref{EQDIF} with initial conditions
\begin{align}
  \phi_k(\eta_0)&=\frac1{\sqrt{\omega_k(\eta_0)}},
  &
  \phi'_k(\eta_0)&=i\omega_k(\eta_0)\phi_k(\eta_0)~.
  \label{GSconditions}
\end{align}
From now on we call $\phi_k(\eta,\eta_0)$ the solution of \eqref{EQDIF},
\eqref{GSconditions}.
At another conformal time $\eta_1$, since the vacuum satisfies different initial
conditions, the theory has a different particle content described 
by 
new creation and annihilation operators $\hat{b}_{\vec k}^\dagger, \hat{b}_{\vec k}$. 
The mode operators at $\hat{a}_{\vec k}$ and $\hat{b}_{\vec k}$ are related by the 
Bogoliubov transformation
\begin{align}
  \hat{b}_{\vec k} = \alpha_k \hat{a}_{\vec k} + \beta_k^* \hat{a}_{\vec k}^\dagger~,
  \label{bogo}
\end{align}
where $\alpha_k$ and $\beta_k$ are independent on $\eta$ and defined by
\begin{align}
  \alpha_k &= \frac{W[\phi_k(\eta,\eta_1), \phi_k(\eta,\eta_0)^*]}{2i}~,
  \\
  \beta_k^* &= \frac{W[\phi_k(\eta,\eta_1), \phi_k(\eta,\eta_0)]}{2i}~,
  \label{bogocoeff}
\end{align}
being $W[x,y]=(\partial_\eta x)y-x(\partial_\eta y)$ the Wronskian.
The new vacuum is related to the old one via
\begin{align}
  \ket{\Omega(\eta_1)} = \prod_{\vec k} \frac1{|\alpha_k|^{1/2}} \exp
  \left(-\frac{\beta^*_k}{2\alpha_k}\hat{a}_k^\da \hat{a}_{-k}^\da\right) \ket{\Omega
  (\eta_0)}~.
  \label{OmegaEta0Eta}
\end{align}
When $\eta_0=n\Pi$  and $\eta_1=(n+1)\Pi$ the above 
  transformation defines the operator  $\mathcal U_n$ which maps the 
state of the $n$th cycle to the next one.
In the following sections we study the vacuum in a bouncing universe,
discussing the possible creation of particles and the occurrence of a
geometric phase after one or more periods. 

\section{Particle production}
\label{partprod}

In a bouncing universe the scale factor of the universe oscillates.
If one assumes a conformally coupled scalar field, the effect of
gravity is included in the $\eta$-dependent frequency $\omega_k(\eta)$ through the mass of the scalar field, 
which is periodic with a certain conformal period $\Pi$. As it can be expected, in general
the particle content at conformal times $\eta$ and $\eta+\Pi$ is the not the
same.  In this section we show that the low-energy vacua at different periods are
not the same in general, i.e. $\ket{\Omega(\eta{+}\Pi)}\neq\ket{\Omega(\eta)}$, 
even though the functions $\phi_k(\eta,\eta_0)$ and $\phi_k(\eta,\eta_0+\Pi)$ 
satisfies the same differential equation \eqref{EQDIF}.

We consider the universe in its ground state at conformal time $\eta_0$. 
As discussed before, an observer at a different
conformal time $\eta_1$ sees different particles which are related to the original
ones by the Bogoliubov transformation \eqref{bogo}. For instance let us call
$a$-particles the particles created by the operators $\hat{a}_k^\dagger$ 
at conformal time $\eta_0$ and 
$b$-particles the ones created by $\hat{b}_k^\dagger$ at conformal time $\eta_1$. 
The vacuum state at $\eta_1$ is the one annihilated by $\hat{b}_k$ so, by definition,
the number of $b$-particles in $\ket{\Omega(\eta_1)}$ is exactly zero.
Nonetheless, $\ket{\Omega(\eta_1)}$ contains many pairs of $a$-particles and
the average density of $a$-particles in  $\ket{\Omega(\eta_1)}$ is
\cite{winitzky}:
\begin{align}
  n_k &= |\beta_k|^2~, & 
\beta_k &= -\frac{\phi'_k
-i\omega_k \phi_k}{2i \sqrt{\omega_k}}~,
  \label{Nk}
\end{align}
where in the second equality we used \eqref{bogocoeff}. 
  For instance, for sudden transitions between two instantaneous vacua,
  the Bogoliubov coefficients are given by
  \begin{align}
    \alpha_k &= \frac12\left(\sqrt{\frac{\omega_k(\eta_1)}{\omega_k(\eta_0)}}
    +\sqrt{\frac{\omega_k(\eta_0)}{\omega_k(\eta_1)}}\right),
    \\
    \beta_k &= \frac12\left(\sqrt{\frac{\omega_k(\eta_1)}{\omega_k(\eta_0)}}
    -\sqrt{\frac{\omega_k(\eta_0)}{\omega_k(\eta_1)}}\right).
    \label{SuddenTrans}
  \end{align}
Therefore if the universe is cyclic and performs a sudden transition from
$\eta_0$ to $\eta_1=\eta_0+\Pi$, then $\beta_k=0$ so the particle content is not 
changed.
On the other hand, when the universe evolves smoothly the Bogoliubov
coefficients are obtained by means of \eqref{bogocoeff} from the solution 
of the differential equation \eqref{EQDIF}. 

The equation \eqref{EQDIF} is known in the literature as the Hill
equation and has many interesting properties \cite{Hills}. In particular, 
because of Floquet theorem, all the solutions of \eqref{EQDIF} can be written as
\begin{align}
  \phi_k(\eta) = e^{i\mu_k\eta} c_k^+ f_k^-(\eta) + 
   e^{-i\mu_k\eta} c_k^- f_k^-(\eta), 
   \label{Floquet}
\end{align}
where the functions
$f_k^{\pm}(\eta)$ are periodic, $f_k^{\pm}(\eta+\Pi)=f_k^{\pm}(\eta)$, 
and $\mu_k$ are called the Floquet exponents.  The 
coefficients $c_k^{\pm}$ describing the mode function are readily
obtained by imposing \eqref{GSconditions}, whereas $\mu_k$ depends 
on $\omega_k$ and, in general, can be real, complex or integer. 
When $\mu_k$ is complex the solutions are not stable, i.e. they diverge for
$\eta\to\infty$. We argue that this is not a physical occurrence so we skip the
discussion of this eventuality. More interestingly, when $\mu_k$ is an integer
(or a fraction) the solution \eqref{Floquet} is periodic of period $\Pi$
(or a multiple of $\Pi$). In this case, 
not only the geometry, but also the quantum fields 
display a cyclic dynamics during the evolution,  so 
$\ket{\Omega(\eta{+}\Pi)}=\ket{\Omega(\eta)}$. 
From the mathematical point of view, 
$\omega_k(\eta)$ can be designed to allow for periodic solutions 
\cite{HillTipo} for each value of $k$, but typically, for physical conformal
factors, the periodic solutions appear only for specific
discrete values of the parameters, e.g. for some specific values of $k$.

In general $\mu_k$ is a real number, making the solution
Eq.\eqref{Floquet},
and thus the quantum fields, an aperiodic function of $\eta$. 
When the vacuum state is not periodic the average density of particles 
after $T$ periods is evaluated from Eqs. \eqref{Nk} and \eqref{Floquet}:
\begin{align}
  n_k(\Pi)&= \frac{\sin (T \Pi  \mu_k )^2}{\omega^2\left|g_k^++g_k^-\right|^2},
  \label{NKpi}
  \\ g_k^{\pm}&= \frac{f_k^\pm(0)}{[\mu_k\mp\omega_k(0)]
    f_k^\pm(0)\mp if_k^{\pm}{ }'(0)}~.
\end{align}
The Floquet exponent $\mu_k$ can be evaluated perturbatively using the
techniques reported in the Appendix \ref{SectFluquet} while the quantities 
$g^\pm$ depends on the initial value. Therefore, $n_k(\Pi)$ is zero only when 
$\mu_k$ is an integer or when $(g_k^\pm)^{-1}=0$. We will discuss below 
these possibilities for some particular cases.

\subsection{Kronig-Penney approximation}
 The equation for the modes of a bouncing universe are formally similar to those
 of Bloch electrons in a spatially periodical lattice, where the space dimension
 is the conformal time. Instead of $\sqrt{\frac{2m}{\hbar}(E-V)}$  for
 $\omega_k$, we now have a relativistic energy given by $\sqrt{k^2+m^2 a(\eta)^2}$, with the periodicity encoded within $a(\eta)$. One can thus use a similar approach to those of used for tunneling of electrons from one lattice site to the other, or for calculating the bands of a solid.
 In this section we will assume for simplicity a scale factor which oscillates as:
\begin{equation}
a(t)=a_0+b\ \sin(r t).
\label{eq:oscill}
\end{equation}
  As $r$ is merely a redefinition of time, we consider $r=1$ hereon. We note
  that this toy model requires $a_0>b>0$ for a regular cosmology in which the
  scale factor never reaches zero. For this model, 
we find that the conformal time 
$\eta=\int^t \frac{d\xi}{a(\xi)}$ is
\begin{eqnarray}
\eta(t)&=&  \frac{2 \tan ^{-1}\left(\frac{a_0 \tan
   \left(\frac{t}{2}\right)+b}{\sqrt{a_0^2-b^2}}\right)}{\sqrt{a_0^2-b^2}} \nonumber \\
&+&\frac{2 \pi  \left(\theta (t-\pi )-\left\|\frac{b}{2}-\frac{t}{2 \pi
   }\right\|\right)}{\sqrt{a_0^2-b^2}}
\end{eqnarray}
where $\|\cdot\|$ refers to the integer part, 
which is everywhere continuous and monotonic, allowing for $\eta(t)$ to be inverted $\forall t$.  From this we obtain a closed
formula for $a(\eta)\equiv a(t(\eta))$ given by:
\begin{eqnarray}
a(\eta) &=&b\ \sin (2 \tan ^{-1}(\frac{\Xi \tan (\frac{1}{2} \eta
\Xi)-b}{a_0}))+a_0
\end{eqnarray}
with $\Xi=\sqrt{a_0^2-b^2} $. 
It is easy to show that the cyclicity of this universe depends on the parameters
$a_0$ and $b$, $a(\eta)=a(\eta+\Pi)$ , with $\Pi=\frac{2\pi}{\sqrt{a_0^2-b^2}}$.
From now on, we will assume $a_0>b$. 
 
\begin{figure}
\centering
\includegraphics[scale=0.6]{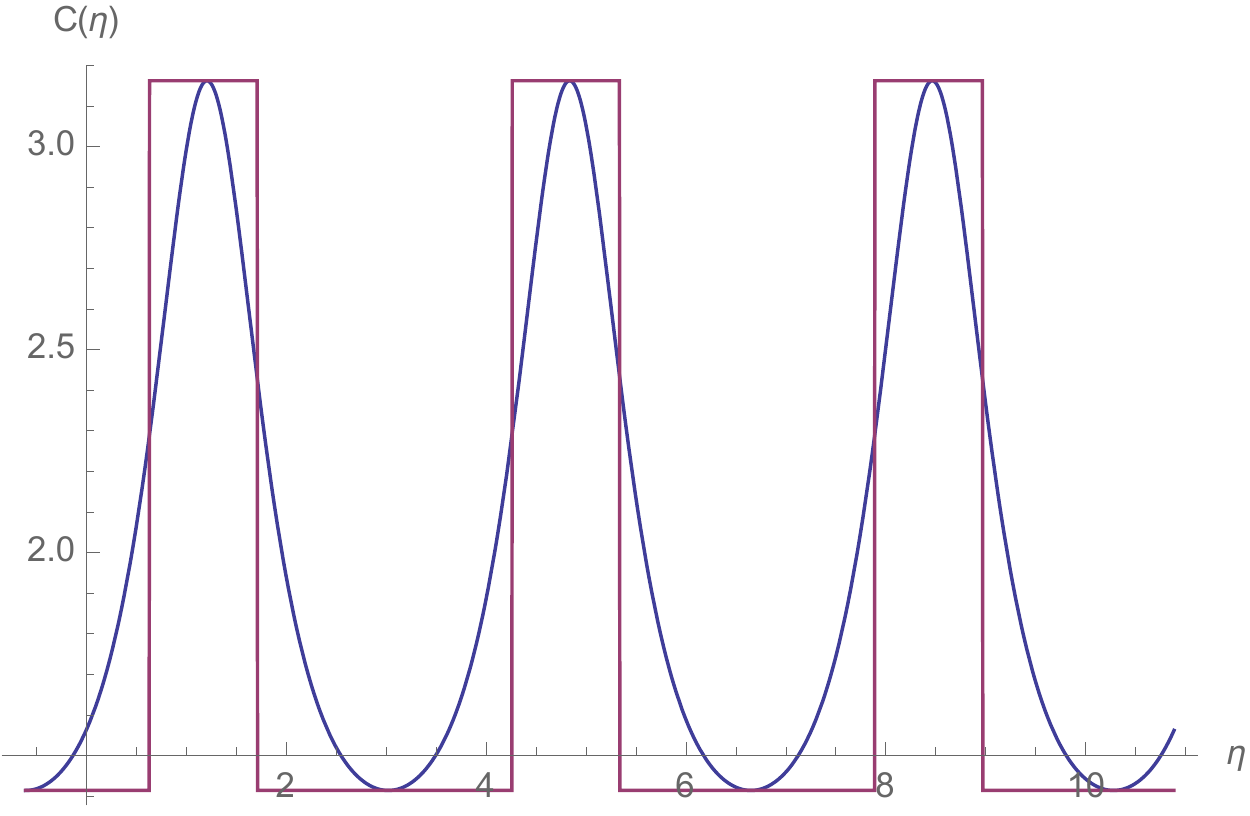}
\caption{Kronig-Penney approximation of the energy $\omega_k(\eta)$ for $k=m=1$. We plot both the real and approximated versions of the effective potential.}
\label{fig:KronigPenney}
\end{figure}

In what follows, we will assume also that our potential can be approximated by step functions. In particular, we would like to treat the effective potential $\omega(k)$ as a a 2-valued function, as in the Konig-Penney approximation for electrons on a Bloch lattice. 
In the standard case of a scalar field in a potential with many local minima, as shown in \cite{Coleman1, Coleman2} the field can move through tunnelings from one local minimum to another. 
In our case instead, the interpretation is different, as the potential is now driven by an external factor, i.e. the bounce of the Universe. 
This will allow to treat the modes close to the bouncing and contracting phases
of this toy Universe as a constant, and thus the local solution being a plane
waves. An example of this approximation is given in Fig. \ref{fig:KronigPenney}.
We use a full width at half maximum (FWHM)  approach, i.e., the energy is
approximated by $\omega_l (k)=\sqrt{k^2+m^2 a^2(\eta_{bounce}))}$ and $\omega_u
(k)=\sqrt{k^2+m^2 a^2(\eta_{contr})}$ in the low and high energy regimes, with $\eta_{bounce}$ being the local bounce, meanwhile $\eta_{contr}$ being the local time of the maximum expansion, and the time of the switch at half maximum, $\eta_{inf}(k)$ and $\eta_{sup}(k)$ are defined by the two (locally in each cycle) solutions of the equation for $\eta$:
\begin{equation}
\omega_k(\eta)= \frac{1}{2}(\omega_k(\eta_{bounce})+\omega_k(\eta_{contr})),
\end{equation}
in which the smaller solution is $\eta_{inf}$, and the larger given by $\eta_{sup}$, denoting the transition from the bounce to the maximum expansion and viceversa, respectively. We denote the $n$-th recurrence of the $\eta_{inf/sup}(k)$ as $\eta^n_{inf}(k)$, as for a lattice. 

For each mode $k$, we solve the contact equation at each transition point, from one bounce to the other one, imposing the continuity and differentiability of the solution. If $\phi_k^n$ is the solution within at  the $n$-th bounce, and $\phi_{k}^{n+1}$ at the $n+1$-th bounce, one can use a transfer matrix approach for calculating the transition coefficients. If $\tilde \phi^{n}_k$ is the wavefunction in the contracting phase, one can write the contact equations between the various solutions at $\eta^n_{inf/sup}$:
\begin{eqnarray}
\phi_k^n(\eta^n_{inf})&=&\tilde \phi_k^n(\eta^n_{inf}) \nonumber \\
\partial_\eta \phi_k^n(\eta^n_{inf})&=& \partial_\eta \tilde \phi_k^n(\eta^n_{inf}) \nonumber \\
\tilde \phi_k^n(\eta^n_{sup})&=& \phi_k^{n+1}(\eta^n_{sup}) \nonumber \\
\partial_\eta \tilde \phi_k^n(\eta^n_{sup})&=& \partial_\eta \phi_k^{n+1}(\eta^n_{sup})  
\label{eq:contact}
\end{eqnarray}
and if one expands the various terms in Fourier modes:
\begin{eqnarray}
\phi_k^n(\eta)= A_n e^{i \omega_l (k)\eta}+B_n e^{-i \omega_l (k) \eta} \nonumber \\
\tilde \phi_k^n(\eta)= C_n e^{i \omega_u (k) \eta}+D_n e^{-i \omega_u (k) \eta}
\end{eqnarray}
as a result of eqns (\ref{eq:contact}), one can write a matricial equation:

\begin{equation}
\begin{pmatrix}
A_{n+1} \\
B_{n+1}
\end{pmatrix}=
T\begin{pmatrix}
A_n \\
B_n
\end{pmatrix}.
\label{eq:transfer}
\end{equation}
In general, the transfer matrix is a rotation due to Bloch theorem, $det(T)=1$, and in particular each element is the wronskian between the two solutions:
\begin{equation}
T=
\begin{pmatrix}
T_{11} & T_{12} \\
T_{21} & T_{22}
\end{pmatrix}=
\begin{pmatrix}
W[\phi_n, \phi_{n+1}] & W[\phi_n, \phi_{n+1}^\dagger] \\
W[\phi_n^\dagger, \phi_{n+1}] & W[\phi_n^\dagger, \phi_{n+1}^\dagger]
\end{pmatrix}.
\label{eq:wronskiantransfer}
\end{equation}
Thus, the element $T_{12}$ represents $2i$ times the Bogoliubov coefficient
which represents $\beta_k$, connecting the modes from one bounce to the next.
Due to the symmetry of the problem, the transfer matrix depends only on the mode
$k$, once one has fixed the mass, and the \textit{shape} of the function $a^2(\eta)$.
From the physical point of view, we are calculating the tunneling of the modes between one bounce to the other, i.e. from one ground state to the other in this approximation.

The elements of the transfer matrix derived in the Kronig-Penney approximation
of the cosmological bounce. We define 
  $\Delta \eta=\eta_{sup}-\eta_{inf}$ and $\overline \eta=\eta_{inf}+\eta_{sup}$. The transfer matrix elements are given by $T_{ij}$, with $1\leq i,j\leq 2$. Thus we have:

\begin{eqnarray}
T_{11}&=&\frac{e^{-i \omega_l(k) \Delta \eta}}{2 \omega_l(k) \omega_u(k)}[2  \omega_l(k)  \omega_u(k) \cos( \omega_l(k) \Delta \eta) \nonumber \\
&+&i ( {\omega_l(k)}^2+{\omega_u(k)}^2) \sin(\omega_u(k) \Delta \eta)], \nonumber \\
\nonumber \\
T_{12}&=& \nonumber \\
-T_{21}&=&i m^2 \frac{a^2(\eta_{contr})-a^2(\eta_{bounce})}{2 \omega_u(k) \omega_l(k)}  \nonumber \\
&\cdot& e^{-i (\omega_u(k)+\omega_l(k)) \overline \eta}
[e^{2 i \omega_u(k) \eta_{inf}}-e^{2 i \omega_u(k) \eta_{sup}}], \nonumber \\
&=&i m^2 \frac{a^2(\eta_{contr})-a^2(\eta_{bounce})}{2 \omega_u(k) \omega_l(k)} \nonumber \\
&\cdot& e^{-i (\omega_u(k)+\omega_l(k)) \overline \eta}  e^{2 i \omega_u(k) (\eta_{inf}+\frac{\Delta \eta}{2}) }\sin( \omega_u(k)  \Delta \eta) \nonumber \\
&\propto& m^2 \frac{a^2(\eta_{contr})-a^2(\eta_{bounce})}{2 \omega_u(k) \omega_l(k)} \sin( \omega_u(k)  \Delta \eta), \nonumber \\
T_{22}&=&\frac{1}{4 \omega_l(k) \omega_u(k)} e^{-i[(2 \omega_u(k)+\omega_l(k)) \eta_{inf}+(\omega_u(k)-\omega_l(k))\eta_{sup}]}  \nonumber \\
&\cdot& [e^{i \omega_l(k) \eta_{inf}}(e^{2i \omega_u(k) \eta_{inf}}-e^{2i \omega_u(k) \eta_{sup}}) \nonumber \\
&\cdot& ({\omega_u(k)}^2+{\omega_l(k)}^2)\nonumber \\
&+&4e^{i [(\omega_l(k)+\omega_u(k)) \eta_{inf}+\omega_u(k) \eta_{sup}]} \nonumber \\
&\cdot& \omega_u(k) \omega_l(k) \cos(\omega_u(k) \Delta \eta)], \nonumber 
\end{eqnarray}
which one can verify to satisfy the condition of the transfer matrix: (i)
$det(T)=1$, (ii) for $\Delta \eta\rightarrow0$ or
$\omega_l(k)\rightleftarrows \omega_u(k)$, $T_{11}\rightarrow T_{22}\rightarrow
1$, meanwhile $T_{12}\rightarrow 0$. Since we are interested in the off-diagonal
elements of this matrix, we note that for $m=0$, the Bogoliubov coefficient goes
to zero; in addition to this, it is proportional to the difference between the
maximum and minimum conformal factors, $a^2(\eta_{bounce})-a^2(\eta_{contr})$. It is also clear, that the Bogoliubov coefficient is proportional to $\sin(\omega_u(k) \Delta \eta)$, and thus the faster the Universe expands and contracts, the less particles are produced. 
\begin{figure}
\includegraphics[scale=0.5]{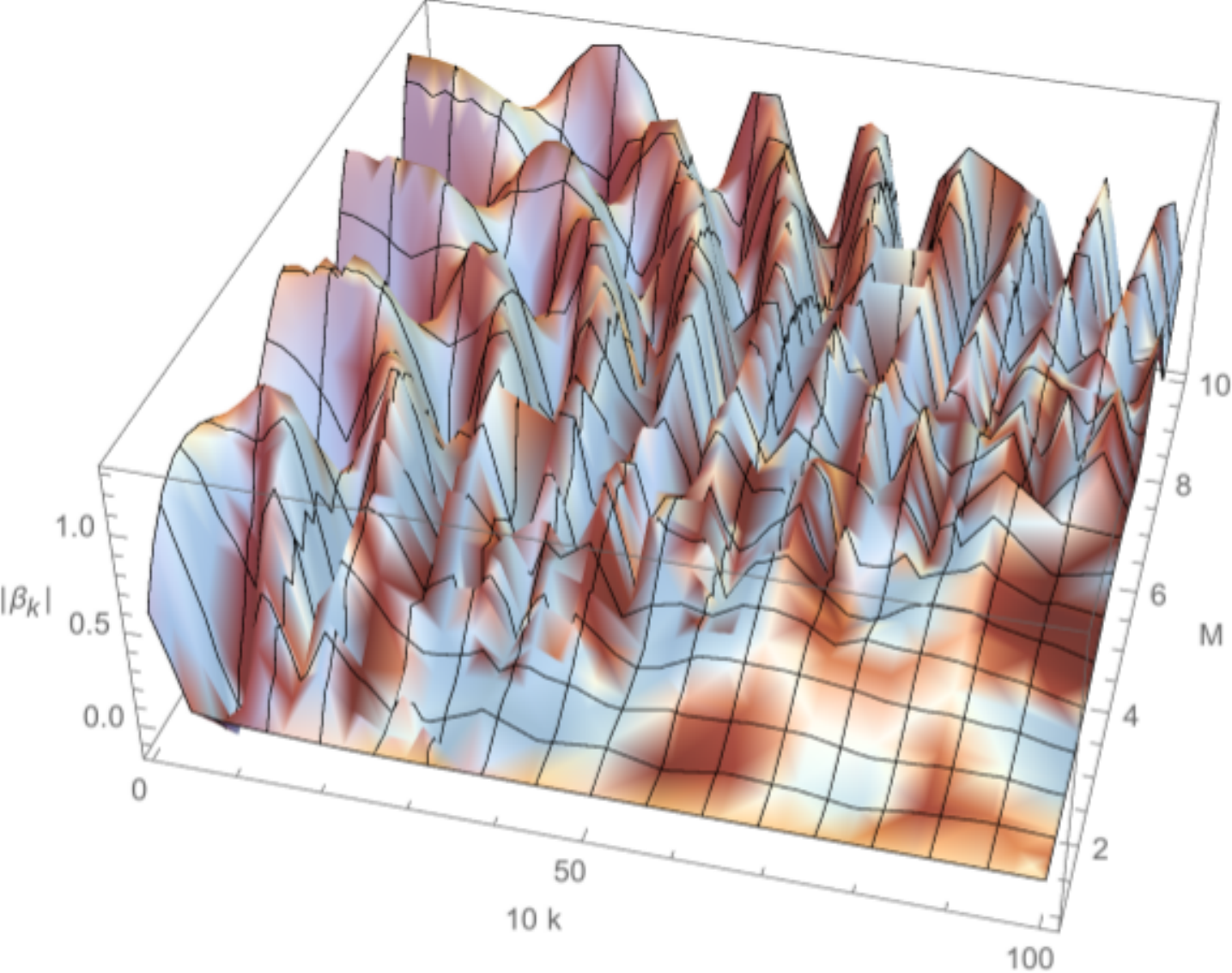}
\caption{In this figure we plot $|\beta_k(m)|)$ as a function of $m$ and $k$. We observe both interference effects and thresholds related to the mass of the scalar field. These features will be also present in the other approximations.}
\label{fig:betakp}
\end{figure}

In Fig. \ref{fig:betakp} we plot the Bogoliubov coefficient between one bounce
and the next in the Kronig-Penney approximation for various values of $k$ and
$m$.
As we will see, many of the phenomenological features of this approximation are retained in the exact solution of the differential equation with a smoother approximation of $a(\eta)$, as for instance interference patterns and particle production which goes to zero as $k\rightarrow \infty$. We note, however, that in this approximation the transfer matrix is independent from the bounce number $n$ which we used to diversify one bounce from the other. One can, in fact, interpret the Kronig-Penney approximation as a 2-way WKB approximation at the Bounce and at the point of maximum expansion in which one has $\partial_\eta \omega_k(\eta)=0$. As a last comment to conclude this section, we note that due to the discontinuous approximation of the potential, several interference effects which are spurious occur. This will be solved in the next section, by means of a smoother approximation.

\subsection{Mathieu fields}
\label{sectionMathieu}
The Kronig-Penney approximation is a first, rough approach to calculating the Bogoliubov coefficients. 
In this section we work on the analytical solution of the differential equation which 
describes the evolution of the modes.
Differently from the case discussed before, we consider a simple smooth approximation 
of the conformal factor given by the function:
\begin{align}
  a^2(\eta) = a_0+b\cos(2\eta)~,
  \label{Csimple}
\end{align}
whose period is $\Pi=\pi$. 
Within this approximation, 
the differential equation \eqref{EQDIF} is the well-known Mathieu equation,
whose solutions are given by \eqref{Floquet} with the following 
analytical expressions 
\begin{align}
e^{\pm i \mu_k\eta} f_k^{\pm}=&
\text{C}\left(a_0 m^2{+}k^2,{-}\frac{1}{2} b m^2,\eta\right)\pm \\\nonumber&i
\text{S}\left(a_0 m^2{+}k^2,{-}\frac{1}{2} b m^2,\eta\right)~,
  \label{Mathieu}
\end{align}
where $\text{C}(\cdot,\cdot,\eta)$ is the even Mathieu function,
$\text{S}(\cdot,\cdot,\eta)$ is the odd Mathieu function and, in this case,
$\mu_k$ the Mathieu exponent \cite{abramowitz}. 
Mathieu functions have a long history of applications in quantum mechanics
(see e.g. Slater \cite{Slater}) with important applications to the
theory of optical lattices \cite{OpticalO,Optical}. 
From \eqref{NKpi} one finds that after $T$ full oscillations the average
number of quasi-particles is 
\begin{align}
  n_k&=\left|\frac{1-g_k^2}{2g_k}\sin(\pi\mu_k T)\right|^2,\\
  g_k&=
  \frac{ \text{C}\left(a_0 m^2{+}k^2,{-}\frac{1}{2} b m^2,0\right)}{
    \text{S}'\left(a_0 m^2{+}k^2,{-}\frac{1}{2} b m^2,0\right)}
\omega_k(0)
    ~.
  \label{NkMathieu}
\end{align}
\begin{figure}[t]
  \centering
  \includegraphics[width=.4\textwidth]{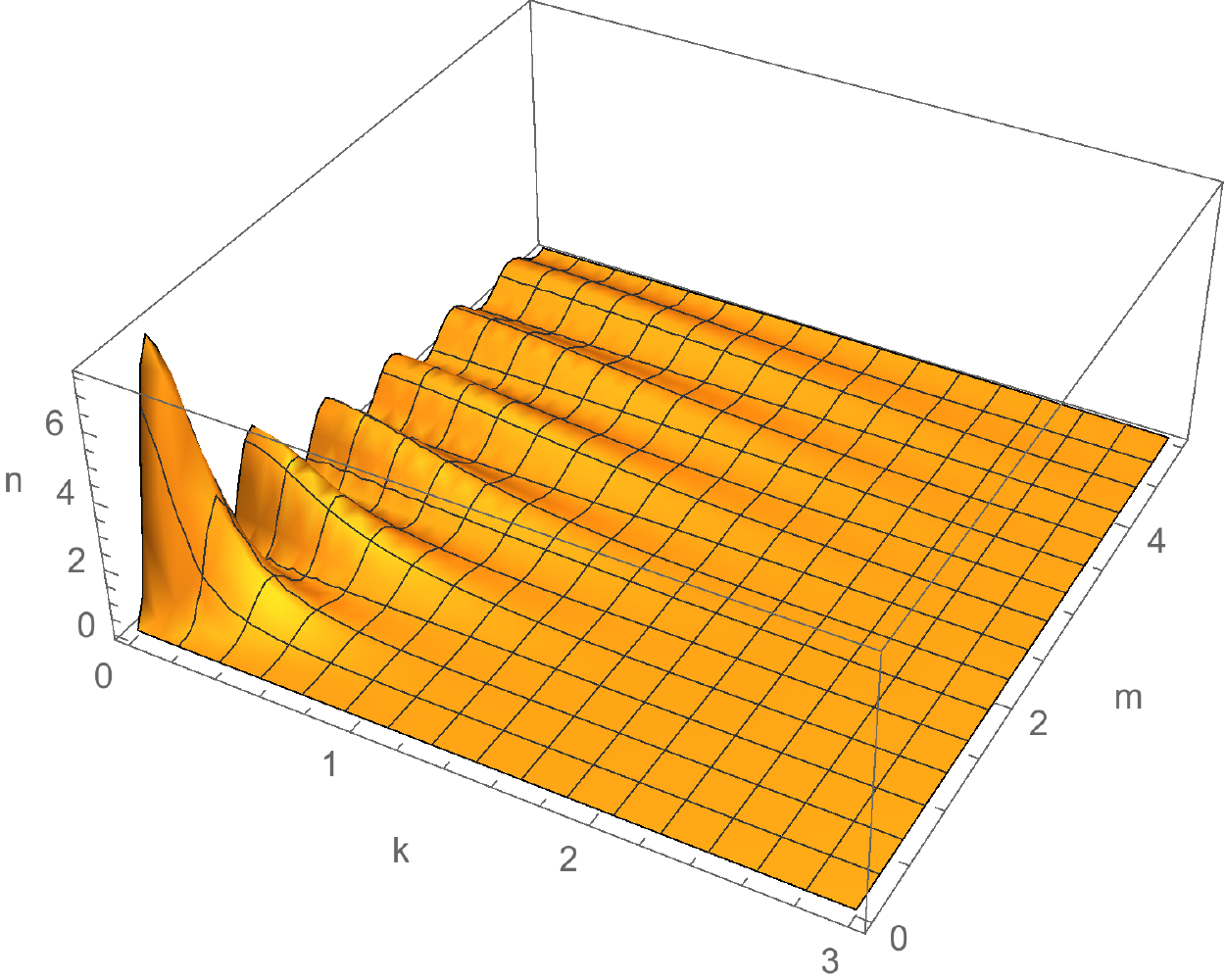}
  \includegraphics[width=.4\textwidth]{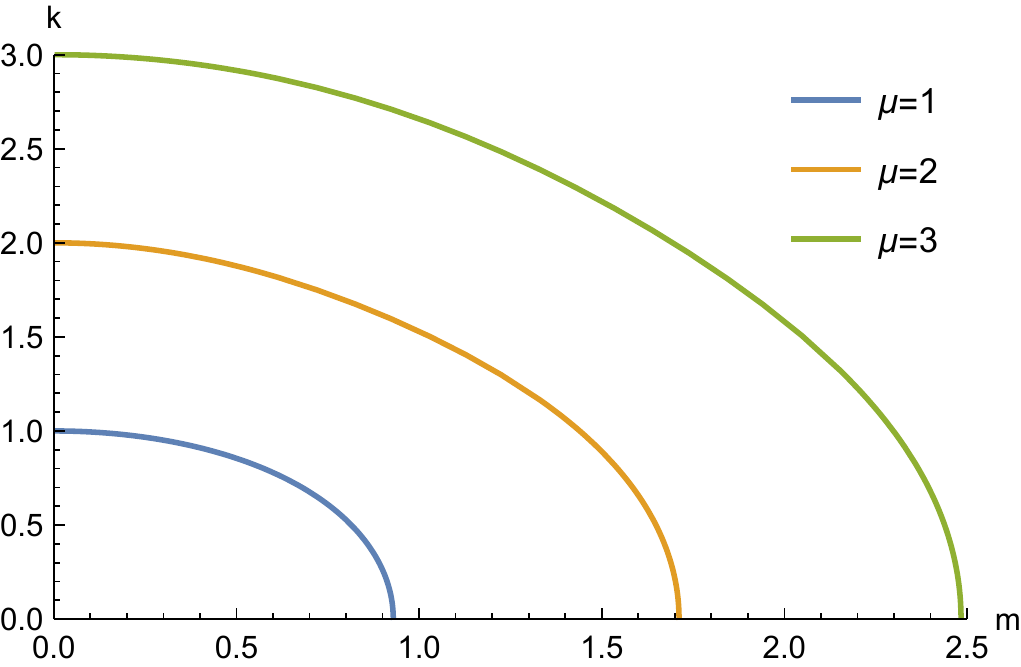}
  \caption{(top) 
    Average number of particles $n$ after a full oscillation ($T=1$) in terms 
    of momentum $k$ and mass $k$, given by 
    Eq.~\eqref{NkMathieu}. (bottom) Value of $(k,m)$ such that $\mu_k$ is 
    integer. We used $a_0=2.0, b=-1.9$.
  }
  \label{figNkMathieu}
\end{figure}
For the present analysis we are interested only in the regions where $\mu$ 
is real but, for the sake of completeness, it is worth mentioning that 
there are some zones in the parameter space where the Mathieu exponent
can become complex and the solutions of eqn. \eqref{EQDIF} may grow 
exponentially as $T\to\infty$ \cite{Enghi}.
From Fig.~\ref{figNkMathieu}(a) one can see that, for small values of  $k$, the energy gap
$\Delta=\min_\eta \omega_k(\eta)$ is sufficiently small to allow for the creation
of some particles. 
Similar interference effects are found also using the Kronig-Penney approximation, 
although emphasized due to the non-smooth approximation in of $a(\eta)$. 
It is interesting to see that, for fixed small $k$, there are some values 
of $m$ where no particles are created. This can be explained by noting 
that for some values of $(k,m)$ the Mathieu exponent $\mu_k$ can be an
integer number, making $n_k=0$ in \eqref{NkMathieu}. The locus of points 
$(k,m)$ where $\mu_k$ is integer is shown in Fig.\ref{figNkMathieu}(bottom).

On the other hand, when $a_0\gg b,m$ we expect that the system remains in its ground
state during the evolution because of the large energy gap $\Delta$ which
prevents the creation of particles. In more mathematical terms, indeed we find
that in this limit $g_k\approx 1$ making $n_k\simeq 0$. 

An interesting question is whether the number of particles remains finite after
many oscillations. Indeed, if 
the current lifetime of the universe is much larger than the oscillation
time, then in our universe we can expect to observe extra particles due to the
oscillating geometry. Calling $N=\sum_{\vec k} n_{k} = 4 \pi \int dk\,k^2 n_k$
the total number of particles 
we find
\begin{align}
  N = \frac{\pi}{2} \int_0^\infty dk \left(\frac{1-g_k^2}{g_k}\right)^2 k^2 +
  \mathcal O(T^{-1/2})~,
  \label{e.NTlarge}
\end{align}
where the $\mathcal O(T^{-1/2})$ correction is due to the stationary phase
approximation and can be neglected for $T\gg1$. 
In our numerical simulations $N$ is finite for any reasonable value of 
$a_0$ and $b$.  Moreover, for large energy gaps, namely when  $a_0\gg b,m$ we
still 
find that $N\approx 0$, showing that the particle content of the universe 
is cyclic. 

We have discussed some limits where not only the geometry but also the
quantum fields show a cyclic or almost cyclic evolution. In the next section
we ask whether this peculiarity shows some quantum features in the form of a 
non-zero geometric phase. 

\section{Berry's phase}
\label{berryphase}
The Berry phase is a geometrical phase acquired by a quantum mechanical system 
during its evolution. It was introduced in 
\cite{BerryAdia} for adiabatic evolutions and then promptly extended to more
general cases \cite{BerryNonadia}.
This geometrical phase is gauge-invariant and can be regarded as a generalization of the 
Aharonov-Bohm phase, which on the other hand appears only in specific cases 
when the fields are coupled to electromagnetic radiation. 
Aside from its fundamental implications, the Berry phase has been used also to
gain further understanding of complicated phenomena such as quantum phase transitions
\cite{QPT1,ZhuQPT} and the dynamics of entanglement \cite{Dario,Entanglement},
while its non-abelian generalization provides an alternative avenue for 
quantum computation \cite{Zanardi}.

%
In a cyclic universe one can interpret the occurrence of a Berry phase as a
memory of the quantum system on the previous cycle. 
From this perspective, we are indeed interested in showing that for the case 
of cyclic cosmologies, such phases could occur to be non-zero. 
As a matter of fact, assuming that the Universe is topologically equivalent to a sphere in which the radius for the spectator scalar field represents an external forcing
undergoing a loop in the space of parameters. Thus, the conditions for the emergence of a non trivial phase are present.
Moreover, we have previously discussed that, if a Universe is cyclic, the vacuum states
can display a periodic evolution or almost periodic evolution, depending on the
geometry of the space-time. For cyclic evolutions, thus, a non-zero Berry phase can emerge in the ground state. 

The Berry phase $\phi_B$ for a state $\ket{\Omega(\eta)}$ which undergoes a
cyclic evolution of period $\Pi$ is given by 
\cite{BerryNonadia} 
\begin{align}
  \phi_{\rm B} = i\int_0^\Pi
  \langle{\Omega(\eta)}|\frac{\partial}{\partial\eta}{\Omega(\eta)}\rangle\,d\eta~. 
  \label{BerryPhase}
\end{align}
We specify to the case in which $\ket{\Omega(\eta)}$ is the ground state of the 
universe at conformal time $\eta$, and we assume that the universe was in its
ground state at time $\eta=0$. Exploiting the relation \eqref{OmegaEta0Eta} between
ground states at different conformal times one can show that the Berry
connection is 
\begin{align}
  \inner{\Omega}{\Omega'} &= \frac12 \sum_{\vec k} \left(-\frac{|\alpha_k|'}{|\alpha_k|}
  -\frac{\alpha_k'}{\alpha_k} |\beta_k|^2 + \beta_k^*{}'\beta_k \right)
  \\&=\sum_{\vec k}\frac14\left(\frac{\alpha_k'}{\alpha_k}-
  \frac{\alpha_k^*{}'}{\alpha_k^*}\right) + \frac{\beta_k^*{}'\beta_k-
\alpha_k^*\alpha_k'}{2}~.
  \label{DerivOmega}
\end{align}
The above equation can be written in a more convenient form by defining 
the quantity 
\begin{align}
  \gamma_k = \frac{v_k}{v_k'+i \omega v_k}~,
  \label{GammaK}
\end{align}
which satisfies the differential equation
\begin{align}
  \gamma_k' = 1-2i\omega_k \gamma_k -i\omega_k' \gamma_k^2~,
  \label{GammaPrime}
\end{align}
with the initial condition $\gamma_k(0) = [2i\omega(0)]^{-1}$.
Indeed, by substituting \eqref{bogocoeff} in \eqref{DerivOmega} one finds
\begin{align}
  \phi_{\rm B} &= 
  \int_0^\Pi 
  \sum_{\vec k}\frac{\omega'_k(\eta)}{2} \Re[\gamma_k(\eta)]\,d\eta~,
  \\ &=2\pi
  \int_0^\Pi d\eta \int_0^\infty dk\,
  k^2\omega'_k(\eta) \Re[\gamma_k(\eta)]~.
  \label{BerryIntegral}
\end{align}

\begin{figure}[t]
  \centering
  \includegraphics[width=0.45\textwidth]{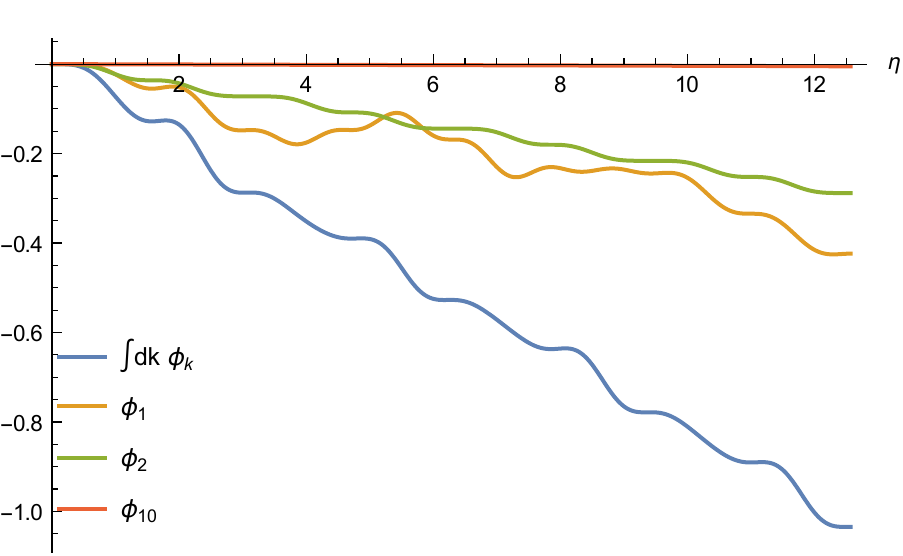}
  \caption{Integral $\int dk \phi_k$ and phases $\phi_k$ for different values of
    $k$ as a function of $\eta$. The conformal factor is given by  
    Eq.\eqref{Csimple} with $a_0=2$, $b=1$, while $m=1$. }
  \label{fig:Phi}
\end{figure}

Before evaluating the Berry phase for some relevant parameters, we 
study the dependence on $k$ of the integrand $I_k = 2\pi k^2\omega'_k(\eta)
\Re[\gamma_k(\eta)]$. When $k\gg m_{\rm eff},m_{\rm eff}'$ it is 
$\omega_k \simeq k$, $\omega_k'\simeq 0$ and hence $I_k\simeq 0$. 
Therefore, one expects that the integral over $k$ in eqn. \eqref{BerryIntegral}
can be performed up to a cut-off value $k_{\rm cut}$ without introducing 
significant errors. In particular, here we evaluate the Berry's phase
numerically by solving the coupled differential equation 
$\phi_k'(\eta) = I_k(\eta)$ 
and eqn. \eqref{GammaPrime} for several values of $k \in[0,k_{\rm
cut}]$ and then using $\phi_B=\int_k \phi_k(\Pi)\,dk$, 
where the latter integral is evaluated numerically. 
For instance, in Fig. \ref{fig:Phi}, we show that $\phi_k \simeq 0$ for $k\gtrsim
10$, while on other hand the integral $\phi(\eta)= \int_k \phi_k(\eta)\,dk$ is non-zero
and increases for increasing $\eta$. 
\begin{figure}[t]
  \centering
  \includegraphics[width=0.45\textwidth]{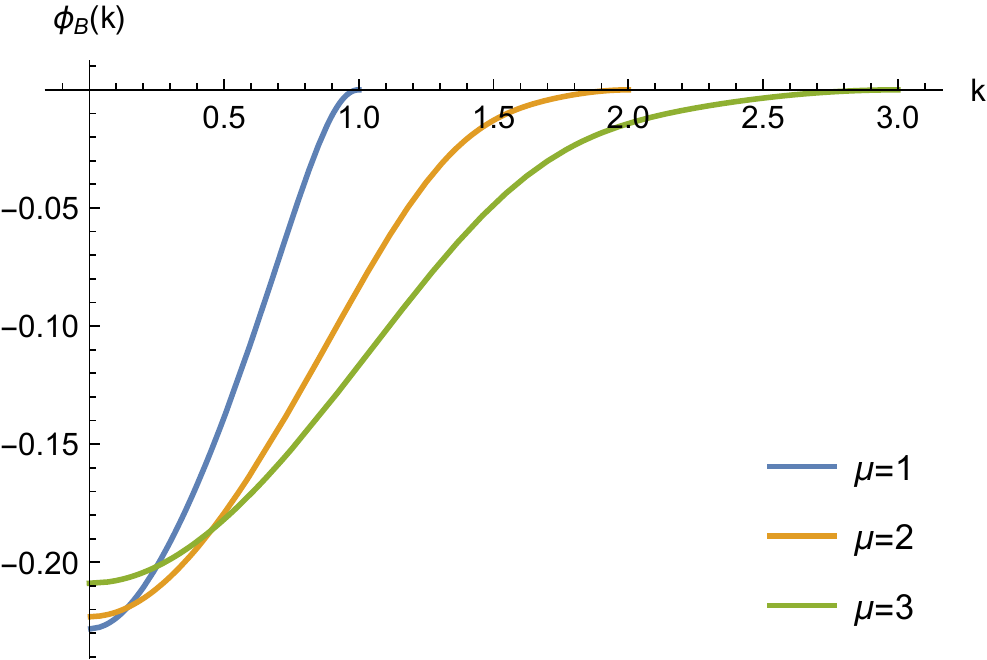}
  \caption{$k$-dependent Berry's phase $\phi_B(k)$ as a function of $k$,
    evaluated along the curves in the $(k,m)$ space 
    shown in Fig.\ref{figNkMathieu}(down).
  }
  \label{fig:Phik}
\end{figure}

From the form of the Berry connection one can see that, if there is no particle
creation during the evolution of the universe, i.e. $\beta_k(\eta)=0$ for each 
value of $\eta$, then the Berry phase is exactly zero. In a non-trivial cyclic 
evolution on the other hand, some particles are created during the bounce, but 
then after one period the fields go back to their initial state. 
This observation, that will be clarified in the following, motivates our
interpretation of the Berry phase as a memory of the dynamics of the universe. 
However, as
in eqn. \eqref{BerryIntegral} the constraint about the periodicity of the initial
state is not explicit, we start our analysis by introducing 
the $k$-dependent Berry phase defined by 
\begin{align}
  \phi_{\rm B}(k) &= 
  \int_0^\Pi 
  \frac{\omega'_k(\eta)}{2} \Re[\gamma_k(\eta)]\,d\eta~.
  \label{BerrySingleMode}
\end{align}
Indeed,
in section \ref{sectionMathieu} we have shown that, for a given cosmological
factor, there are some pairs $(k,m)$ where there is no particle creation
after a bounce. In other terms, particles are created during the evolution of
the universe but after a cycle the ground state is exactly the initial ground 
state. 
In particular, for the simple cosmological factor 
\eqref{Csimple}, we have shown that 
the locus of points where $n_k(\Pi)=0$ define different
curves (see Fig. \ref{figNkMathieu}) 
in the $(k,m)$ space where the Mathieu exponent $\mu$ takes 
integer values. We now evaluate the $k$-dependent Berry phase along those
curves and  show that the Berry's phase is non-zero 
even when the ground state is perfectly periodic, 
thus signaling a non-trivial quantum evolution of the universe. 
The results are shown in  Fig. \ref{fig:Phik}.
Moreover, we observe that $\phi_B(k)$ is larger for smaller values of 
$k$ where the corresponding mass in the curves  Fig. \ref{figNkMathieu} 
is smaller. 

\begin{figure}[t]
  \centering
  \includegraphics[width=.45\textwidth]{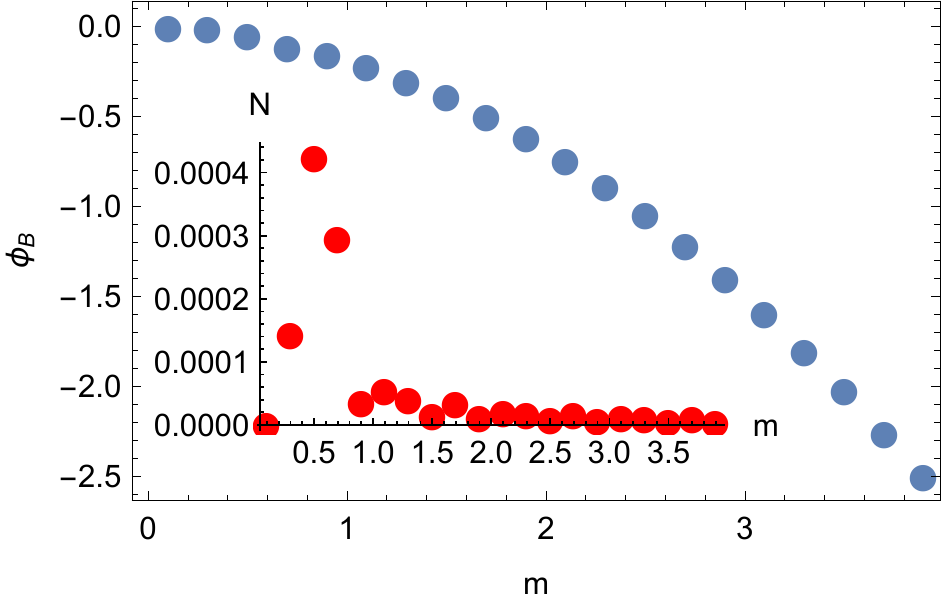}
  \caption{Berry's phase after one cycle $\phi_B$ 
    as a function of the mass $m$.  The oscillating 
    geometry is implemented by the factor factor \eqref{eq:oscill}.
    Inset shows the mean density of particles in the state 
    $\ket{\Omega(\Pi)}$. 
  }
  \label{fig:berry}
\end{figure}

The results presented in  Fig. \ref{fig:Phik} show a rigorous findings of a 
non-zero Berry phase acquired after one cycle by some discrete set of modes for
which $n_k\equiv 0$. However, in the universe there are many modes and, 
as discussed in the section \ref{partprod}, in general the vacuum after 
one period $\ket{\Omega(\eta_0{+}\Pi)}$ is different from
$\ket{\Omega(\eta_0)}$. The closeness of these two vacua can be evaluated with
the fidelity
\begin{align}
\mathcal F &= |\langle{\Omega(\eta_0)}|{\Omega(\eta_0{+}\Pi)}\rangle|^2 =
\prod_{\vec k} |\alpha_k|^{-1}
\\\nonumber &= e^{-\frac12\sum_{\vec k}\log(1+|\beta_k|^2)}
\simeq e^{-N/2}~,
  \label{Fid}
\end{align}
where $N=\sum_{\vec k} n_k = 4\pi \int dk\, k^2 |\beta_k|^2$ 
is the mean density of particles 
and where, in the last equality, we assumed that $|\beta_k|^2\ll 1$.
We claim that, even when $\ket{\Omega(\eta_0{+}\Pi)}$ is different
from $\ket{\Omega(\eta_0)}$, and thus the evolution is not perfectly cyclic,
if $N\ll 1$ the two vacua are almost indistinguishable ($\mathcal F\simeq 1$). 
In that limit we can thus consider the field evolution approximately periodic
and evaluate the Berry phase. 

In Section \ref{partprod} we have shown that, as far as the conformal factor 
\eqref{Csimple} is concerned, the condition $N\ll 1 $ can be
obtained when the gap $\Delta$ is large, i.e. when $a_0\gg b,m$. 
In the inset of Fig. \ref{fig:berry} we show that this holds also for the more
complicated geometry of eqn. \eqref{eq:oscill}. Given that in these conditions the
universe is cyclic, in Fig. \ref{fig:berry} we evaluate 
the resulting Berry phase of eqn. \eqref{BerryIntegral} and we find that it is non-zero.
A non-trivial Berry phase seems thus a general occurrence in cyclic universes. 
{
To motivate this further, we consider more general scale factors with higher
harmonics in the Fourier expansion \eqref{Csimple}, namely we set
$a^2(\eta)=\tilde a^2_{n,\alpha}(\eta) \equiv  a_0
+\sum_{j=1}^n a_j \cos(2\eta j)$ and we choose $a_j=j^{-\alpha}/\sum_{k=1}^n
k^{-\alpha}$ so that the dominant frequency is $2$. 
\begin{figure}[ht]
  \centering
  \includegraphics[width=.45\textwidth]{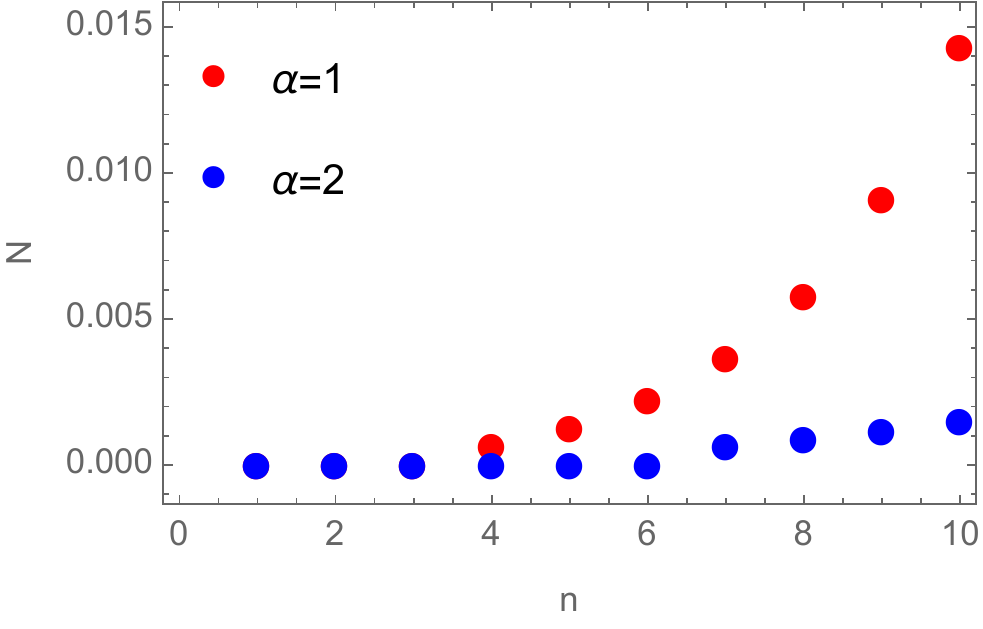}
  \includegraphics[width=.45\textwidth]{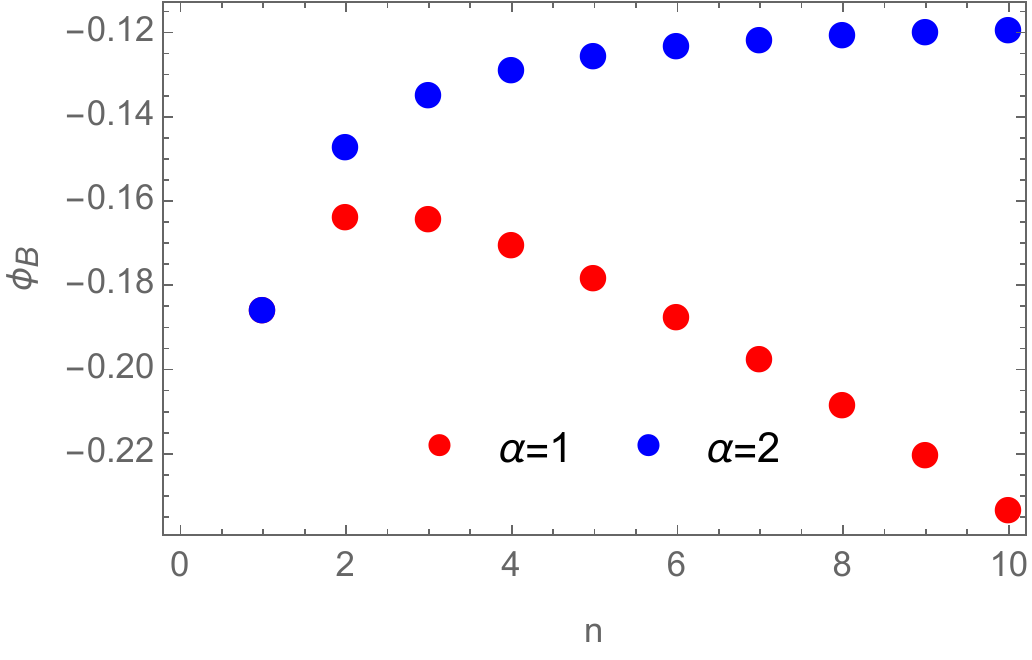}
  \caption{Total number of particles $N$ and Berry's phase $\phi_B$ after the
    first cycle when $a(\eta)=\tilde a_{n,\alpha}(\eta)$ (see discussion in the text) for
  different values of $n$ and $\alpha$. $a_0=20, m=3$. }
  \label{fig:NkPhiB}
\end{figure}
As shown in Fig. \ref{fig:NkPhiB}, when $a(\eta)=\tilde a_{n,\alpha}(\eta)$ the
number of particles increases with $n$, although $N\ll 1$ especially for
$\alpha=2$. In all the cases discussed so far, the Berry phase is non-zero after
a bounce so we argue that it is a generic feature of cyclic universes. 
}

In order to show that the Berry phase heirs information of past evolution of the Universe in this toy model, we consider the accumulated phase in the limit $k^2\gg m^2 a^2(\eta)$. In this case an analytical result can be obtained (see  
Appendix \ref{SectFluquet} for a full derivation), showing that wavelengths at which Berry's phase is zero can be enumerated and given by:
\begin{eqnarray}
k*_{n+} &=&\Pi (2n- \frac{m^2 E}{4 \pi n^2} )
\label{eq:phasezero}
\end{eqnarray}
where $E=\frac{1}{2 \pi} \int_{0}^{\Pi}  a^2(\eta) d\eta$ and $\Pi$ is the length of a cycle measured in conformal time. These wavelengths contain information on the life cycle of the Universe as eqn. (\ref{eq:phasezero}) shows.

  Measuring the Berry's phase requires the introduction of extra degree's of 
  freedom coupled with the field. To show this point, we use a simplified 
  argument and we consider a further term in the potential $V(g)$ in 
  \eqref{e.action} of the form $m_r^2 S^z$, where $S^z$ is a spin-1/2 operator
  in the $z$ direction. Depending on the value of the spin along the $z$
  direction, this extra coupling renormalizes the mass. Because, as we have
  shown, the Berry's phase depends on the mass of the field, after one cycle 
  the state of the coupled system formed by the field and the spin is
  \begin{align}
    \frac1{\sqrt 2}\left(
    e^{i\phi_B^+}\ket{\Omega_+}\ket{S^z{=}{+}\frac12}+
    e^{i \phi_B^-}\ket{\Omega_-}\ket{S^z{=}{-}\frac12}
    \right)
    ~,
  \end{align}
  namely there is a different Berry's phase depending on the state of the spin. 
  The relative Berry's phase can then be measured, e.g., with $\langle
  S^x\rangle$. 
  Although simplified, our argument shows that the Berry's phase can have
  measurable effects when the field is coupled to other external degrees 
  of freedom. 




\section{The case of LQC}

In the preceding sections we have tried to keep the discussion as general as possible. In fact, eqn. (\ref{eq:phasezero}) applies to any scale factor which is periodic in conformal time. In this section, we will calculate the Berry phase using the effective loop quantum cosmology equations for a closed Universe. We start by introducing the standard Friedmann equations in a genericmatter-dominated Universe for $k=1$ in the dust approximation, in which $P=0$ and $\rho(t) a(t)^3 = constant$.
The first Friedmann equation becomes \cite{dodelson}:
\begin{equation}
H=(\frac{\dot a}{a})^2=\frac{8 \pi G}{3} \frac{\rho_0}{a^3}-\frac{1}{a^2},
\label{eq:crunch}
\end{equation}
and if for simplicity we consider the case $A=\frac{8 \pi G \rho_0 }{3}$,
then it is a known fact that the scale factor takes the form $a(\eta)=A(1-\cos(\eta))$. 
Inevitably, such equation describes a Universe in which also a Big Bang is present, as dynamically the scale factor eventually becomes zero at values of conformal time $\eta=\{0,\pi,2\pi,\cdots\,n \pi\}$.

These equations describe a Universe in which there is a maximum expansion (the Big Crunch), followed by a recollapse.
It is a widely accepted idea that during the recollapse phase, when the linear size of the scale factor is of the order of the Planck length, Quantum Gravity phenomena take place \cite{asht1,merc1}. Loop Quantum Cosmology (LQC) is one example in which a Bounce due to quantum corrections occurs. After the bounce, a new expansion phase is recovered, and thus it is possible to observe a cyclic dynamics of the scale factor. Thus, the machinery we have developed can be applied to the case of Loop Quantum Cosmology if considering purely semiclassical Friedmann equations.

\begin{figure}[ht]
  \centering
  \includegraphics[width=.45\textwidth]{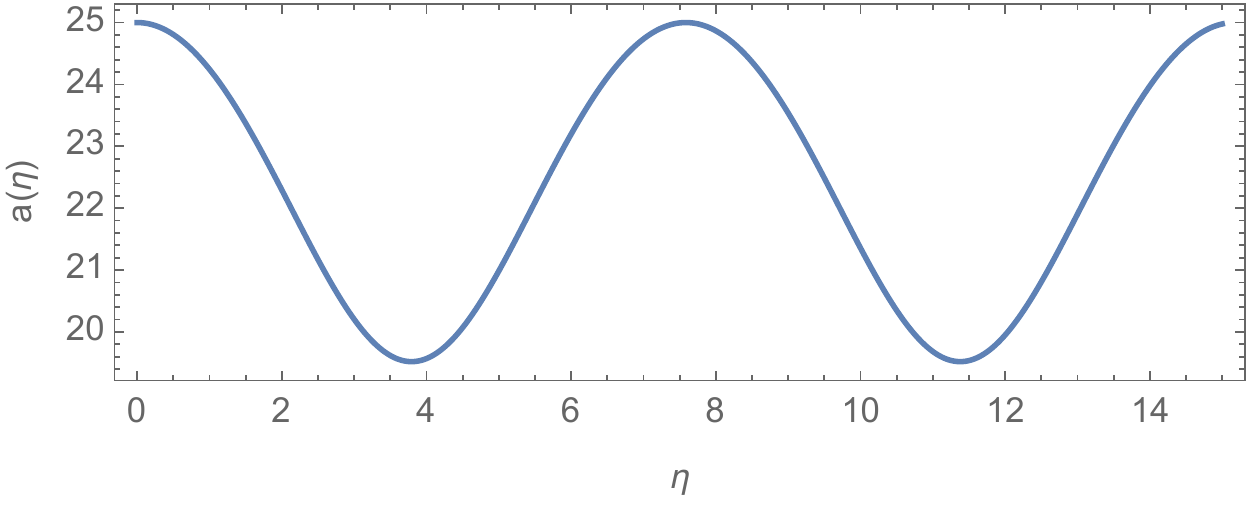}
  \includegraphics[width=.45\textwidth]{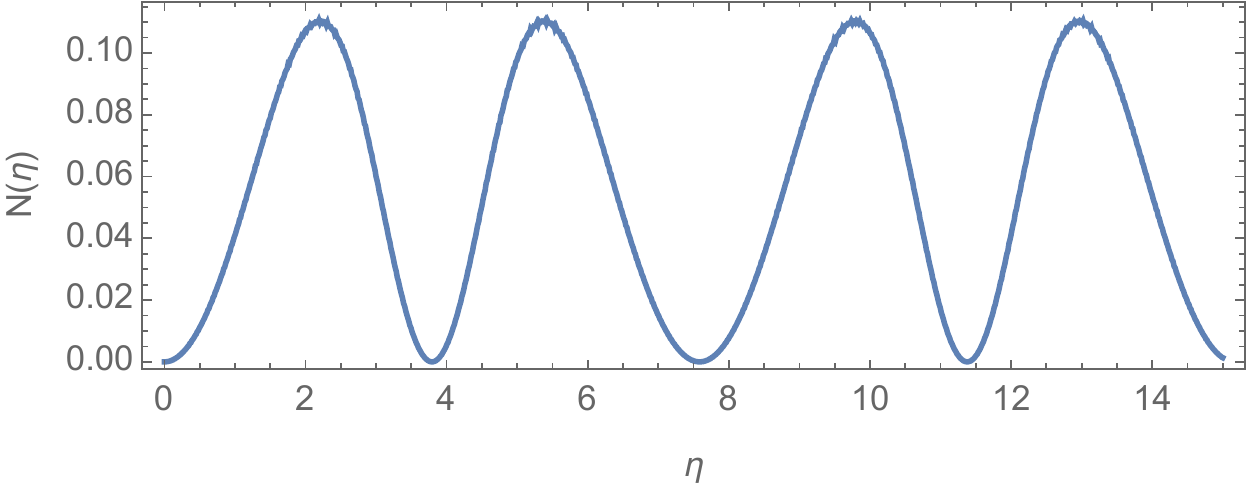}
  \includegraphics[width=.45\textwidth]{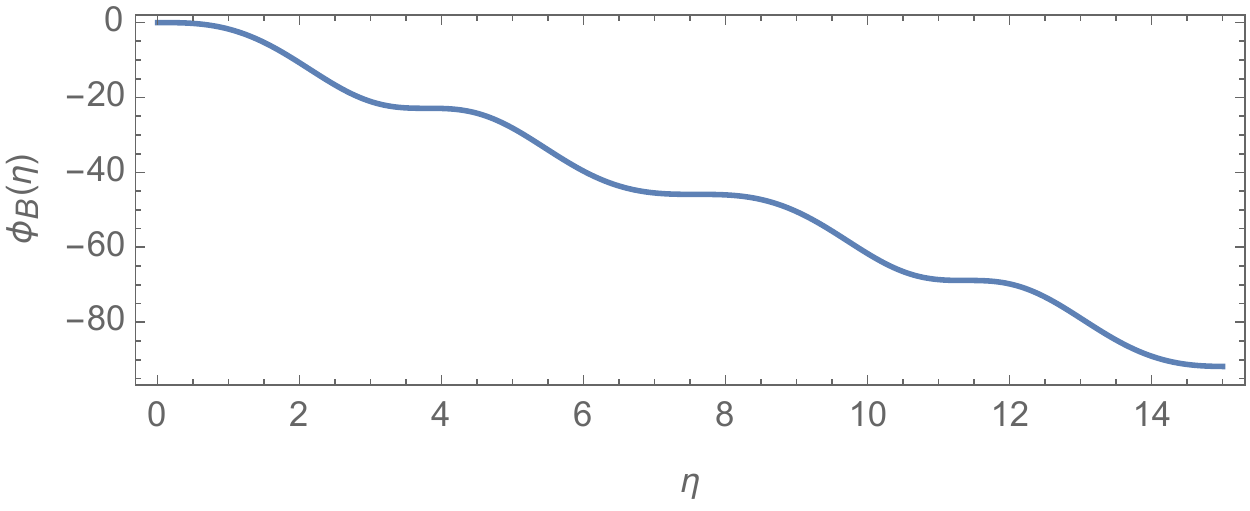}
  \caption{Conformal factor $a(\eta)$, total number of particles $N(\eta)$ and 
    dynamical (Berry) phase $\phi_B(\eta)$ for $m=3, a(0)=25, \rho_{\rm
    LQC}=10^{-4},\rho_0=\rho_{\rm LQC}/2, G=\frac{3}{32 \pi  a(0)^2 \rho_0}$.
  }
  \label{fig:LQC1}
\end{figure}

\begin{figure}[ht]
  \centering
  \includegraphics[width=.45\textwidth]{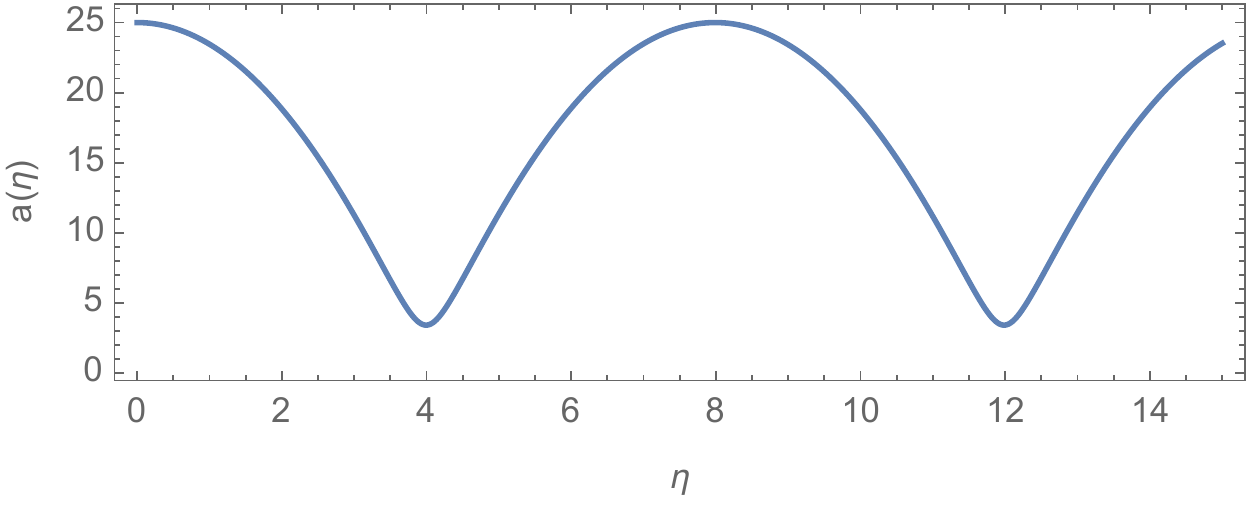}
  \includegraphics[width=.45\textwidth]{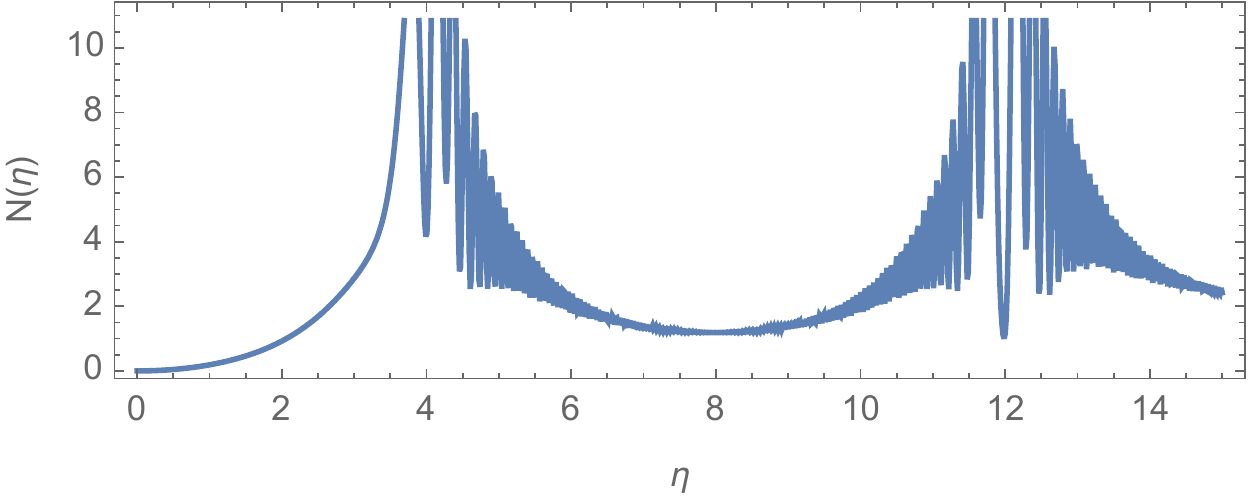}
  \caption{Conformal factor $a(\eta)$ and total number of particles $N(\eta)$ 
    for $m=3, a(0)=25, \rho_{\rm
    LQC}=10^{-4},\rho_0=\rho_{\rm LQC}/100, G=\frac{3}{32 \pi  a(0)^2 \rho_0}$.
  }
  \label{fig:LQC2}
\end{figure}

Here we consider the effective corrected equations derived in a recent paper \cite{ed1}, where eqn. (\ref{eq:crunch}) is corrected by a term which contains a quantum gravitational term
\begin{eqnarray}
(\frac{\dot a}{a})^2&=&\left(\frac{8 \pi G}{3}\frac{\rho_0}{a^3}-\frac{1}{a^2}\right)\left(1-\frac{\rho}{\rho_{LQC}}\right) \nonumber \\
&=&\left(\frac{8 \pi G}{3} \frac{\rho_0}{a^3}-\frac{1}{a^2}\right)\left(1-\frac{\rho_0}{a^3 \rho_{LQC}}\right).
\label{eqn:lqc1}
\end{eqnarray}
In general, equations of the type above strongly depend on the details of the quantization procedure, but the phenomenology described is well captured by eqn. (\ref{eqn:lqc1}), which differs in the exact values of $\rho_{LQC}$.  The effective model of eqn. \ref{eqn:lqc1} agrees with other quantization procedures \cite{cor1} up to higher orders in the quantization procedure, but differs from well known results in the field \cite{asht1}. 
Another important comment to be made here is that
 the derivatives with respect to ``time", are now with respect to an external scalar field which takes the role of a clock, which is a standard procedure in Quantum Gravity approaches \cite{asht1,merc1}. 
 Notwithstanding these important remarks and without going into the merit of the
 quantization procedure, the dynamical interpretation of equation
 (\ref{eqn:lqc1}) is clear: at large scale factor, matter forces the Universe to
 recollapse, meanwhile at scale factors comparable to the Planck length, the
 quantum regime kicks in, forcing the Universe to re-expand. This beautiful and
 appealing picture is the mechanism for which the singularity of the Big Bang is
 avoided, and which we considered in this paper insofar in a stylized manner. In
 order to proceed with our calculations, we need to recast the eqn.
 (\ref{eqn:lqc1}) into a form which is explicitly solvable as a function of the conformal time $\eta$. It is easy to use the definition of conformal time, and turn eqn.  (\ref{eqn:lqc1}) into eqn. (\ref{eq:lqc2}) below:

\begin{eqnarray}
a''(\eta )&=&a^3(\eta ) \Biggl(-\frac{3 A \left(1-\frac{2}{a(\eta )^3 \rho_{LQC}}\right)}{2 a(\eta )^3}\nonumber \\
&+&\frac{A   \left(1-\frac{1}{a(\eta )^3 \rho_{LQC}}\right)}{a(\eta )^3} + \frac{3}{8 a(\eta )^5 \rho_{LQC}}\Biggr)
\label{eq:lqc2}
\end{eqnarray}
for the case $a_0=A=1$. One can in general obtain periodic solutions to the
scale factor, as in Figs. \ref{fig:LQC1},\ref{fig:LQC2}. It is a general
feature, usually discussed in the literature of Loop Quantum Cosmology, that the
Universe remains surprisingly close to the classical trajectories predicted by
eqn. (\ref{eq:lqc2}) also in the quantum regime. 
In this case, we can in fact use the semiclassical method described in the body of this paper, and evaluate the Berry phase numerically after each cycle.

We stress that the Berry phase can be observed only if the dynamics of quantum
wave function of the scalar field is cyclic. This implies that depending on the
parameters $\rho_{LQC}$ and $\rho_0$, we have to check whether the particle
content of the Universe is negligible. In Fig. \ref{fig:LQC1} we plot from the top to the bottom, the scale factor, the particle content and the Berry phase respectively obtained numerically  for the case $\rho_{LQC}=2 \rho_0$, i.e. when the density of the Universe is smaller but comparable to the effective density due to LQC corrections. We observe that the particle content $\ \forall \eta, N(\eta)\ll 1$ due to the smooth Bounce. In this case, we observe a non-zero Berry phase which implies that at the quantum level the scalar field carries information from one Bounce to the other.

In Fig. \ref{fig:LQC2} we show the same plots obtained for the case
$\rho_{LQC}=100 \rho_0$, and thus in the regime in which the Bounce is abrupt.
In this case, we observe a high particle content, which although invalidating
the calculation of the Berry phase, implies that the scalar field carries information through the Bounce in the form of particles. This picture might seem to differ however from the results obtained in (\cite{asht1,woy1,sin1}), where it has been shown that the dynamics has effectively no memory from the previous process. As a first comment, it is worth to note in \cite{barb1} it has been discussed that although similar, the polymeric quantization differs from the standard harmonic oscillator quantization used here. Moreover, our analysis relies on the existence of a spectator scalar field which is quantized in a standard way, and is forced (parametrically) by the dynamics of the scale factor. Thus it is not the Universe \textit{per se} carrying the phase, but the scalar field living on it.

{
Before concluding this section it is worth mentioning that, although in this
paper we focused on standard semi-classical methods, the rigorous origin of the
Berry's phase might be further clarified with a full-quantum treatment, without
any semiclassical approximation, given that the Hilbert space structure is
exactly known \cite{lqcz1}. Specifically, it may be possible to generalize
recent techniques \cite{Dario}, where the transition between quantum to
classical behavior is explicitly discussed, to the case of loop quantum gravity. 
}

\section{Conclusions}

{In the present paper we have studied the quantum features of scalar field both in
cyclic and bouncing toy Universes, and in the case of Loop Quantum Cosmology effective Friedman equations.   Among the many alternatives to the inflaton
field, there are many approaches in which the  Universe is considered cyclic.
Among these, we mentioned Loop Quantum Cosmology and the Ekpyrotic String
approach. Our motivation was to understand whether a scalar field does hold
memory of the previous bounces in subsequent ones, thus changing the physics at
each bounce. 
We studied the particle content after many bounces in a general manner, and observed numerically that even when the 
ground state is periodic, and hence there are no particles after one cycle, 
the universe keeps track of its history via a non-zero Berry phase.
We argue that this is potentially a physically-relevant phenomenon
if the scalar field is entangled, and have provided general arguments why such
phases should be expected for more general, periodic evolutions of the conformal
factor of the Universe. }
As a result of our analysis, we have observed within various approximations, as
for instance the Kronig-Penney and the case of Mathieus approximation, the
presence of particles, and noted that in cases where the particle content
present at each new bounce is negligible, one can reliably observe the presence of a Berry phase.
In fact, 
after having introduced a differential equation describing the Berry's
connection as function of time, we have obtained a numerical solution of the
Berry's phase, showing that this is non-zero for our toy model, but providing a
general framework, based on the Fourier expansion of the energy of the scalar
field, for calculating it perturbatively. 
We have observed that when the ground state is cyclic, 
for a particular mode, one can observe a non-zero Berry's phase. 
In addition to this, we have observed that values for which the Berry phase are zero, contain information on the $L^2$ norm of the scale factor of the Universe over the cycle.
This gauge-invariant phase is of geometrical origin and provides a unified
framework to explain important observable phenomena like the Aharonov-Bohm
effect. Being a measurable  quantity, a non-zero Berry's phase can provide 
a figure of merit to demonstrate that the universe is cyclic. 
This phase
becomes physically relevant if the scalar field is entangled to another field which can be measured. In this case, after each bounce the physical state of the system changes, and thus the physics changes as well.

 {In the first part of this paper, we worked initially} within
 the framework of a toy model and thus there are some further limitations to
 this approach. First of all, we have considered modes which cross the bouncing
 phase without any new physics to intervene, as for instance Quantum Gravity.
 The paper does not describe that phase, and treats the Universe as a smoothly
 oscillating ball. Yet, we believe that this phenomenon is interesting
 \textit{per se} from the theoretical standpoint, {as also for
 generic Fourier expansions of the scale factor a non-zero Berry's phase was
 observed. In the second part of this paper, we applied these ideas to the case
 of Loop Quantum Cosmology (LQC), where a description of the quantum phase of
 these oscillations is provided. Also, an extra appeal of this approach for the
 present paper is that in LQC the quantum description is very close to the
 classical one, and thus we do not require the use of a full quantum
 gravitational approach.  In the case of LQC, we find that for the case of a
 specific semiclassical Friedman equations for which there are oscillations of
 the scale factor, one can observe two specific regimes. In the first regime, in
 which the effective quantum density is comparable to the density of the
 Universe at $a=1$, we observed numerically small particle content and non-zero
 Berry phase. In the other regime, in which the quantum effective density is
 much higher than the density of the Universe, one observes non-negligible particle content. }

{Our results are general enough to be considered in other cases in details. In fact, given the standard quantization procedure of the scalar field, we have considered the case in which the only requirement is that the scale factor has a Fourier expansion. We studied some specific cases, but used known relations for the Hill equation, which underlies the evolution of the modes of the scalar field. .In addition to the remarks above}, nothing prevents this same analysis to be repeated in the case of particles with spin other than zero, which are far more common than scalar particles.
As a final remark, we stress that there are other ways in which a Berry's phase could be originated for non-cyclic cosmologies. This is the case, for instance, if the scalar field undergoes a quantum phase transition in the early Universe \cite{QFU,QPT1,QPT2} and as shown in \cite{bpm} for inflationary models as well.

\section*{Acknowledgments}
We would like to thank Lorenzo Sindoni and Alessandra Gnecchi for comments on an earlier draft of this paper, and to Mercedes Martin-Benito and Edward Wilson-Ewig for helping us and guiding through the results in Loop Quantum Cosmology in a revised version of this article.

L.B. is currently supported by the ERC grant PACOMANEDIA.


\appendix

\section{General considerations and Floquet's theorem}\label{SectFluquet}
The differential equation in eqn. (\ref{EQDIF}),
\begin{equation}
\phi^{\prime \prime} +\omega_k(\eta) \phi=0~,
\end{equation}
is called Hill's equation when $\omega_k(\eta)$ is periodic, and has interesting properties (\cite{Hills}) which can be tackled analytically. 
In particular, according to Floquet's theorem, 
such equation has a pair of solutions which can be written in the form:
\begin{eqnarray}
& \phi_+(\eta)=e^{i \alpha \eta} p_+(\eta) \nonumber \\
& \phi_-(\eta)=e^{-i \alpha \eta} p_-(\eta)~,
\end{eqnarray}
where $\alpha$ is called characteristic exponent and where $p_\pm$  are
functions which have the same periodicity of $\omega_k$. 
If $\alpha$ is not an integer, then one has that 
\begin{eqnarray}
\phi_\pm(\eta+\Pi_k)&=& e^{\pm i \alpha (\eta+\Pi_k)} p_\pm (\eta+\Pi_k) \nonumber \\
&=&e^{\pm i \alpha \Pi_k} e^{\pm i \alpha \eta} p_\pm (\eta) \nonumber \\
&=& e^{\pm i \alpha \Pi_k} p_{\pm}(\eta) \neq \phi_{\pm}(\eta)~.
\end{eqnarray}
In particular, if $\omega_k$ is an even function with period $\Pi_k$, one can expand it in Fourier series:
\begin{equation}
\omega_k(\eta)=\omega^2_{0,k}+ 2 \theta \sum_{n=1}^\infty \gamma_n \cos(2 n
\frac{\eta}{\Pi_k})~.
\end{equation}
$\gamma_n$'s are the expansion factors, and $\theta$ is a factor which we assume
can be considered small. Let us rescale the conformal time such that $\Pi_k=1$, and rescale $\omega_{0,k}$, $\omega_{0,k}\rightarrow \tilde \omega_{0,k}=\omega_{0,k} \Pi_{k}$ and $\tilde \theta=\theta \Pi_k^2$. In this case, then, there is a closed formula for estimating $\alpha$:      
\begin{eqnarray}
\alpha&=&\pm [{\tilde \omega_{0,k}} +\frac{{\tilde \theta}^2}{4 {\tilde \omega_{0,k}}} \sum_{n=1}^\infty \frac{\gamma_n^2}{n^2-{\tilde \omega_{0,k}}^2}+\nonumber \\
&+& \frac{{\tilde \theta}^3}{32 {\tilde \omega_{0,k}}^2} \sum_{n,m= -\infty, n\neq m}^{\infty} \frac{\gamma_{|n|}\gamma_{|m|}\gamma_{|n-m|}}{n m (n+{\tilde \omega_{0,k}})(m+{\tilde \omega_{0,k}})} 
\nonumber \\&+&O(\theta^4)]~.
\end{eqnarray}
For our case of interest, where $\omega_k(\eta)=\sqrt{k^2+m^2 a^2(\eta)}$, we need to relate the coefficients $\gamma_n$ to $\omega_k$.
Assuming the periodicity of $a^2$, one has $a^2(\eta)=\sum_{r=0}^\infty b_r \cos(2 r \frac{\eta}{ \Pi})$;  we now rescale $k\rightarrow \Pi$ and $m\rightarrow m \Pi$, one has that the Fourier coefficients $\gamma_n$ introduced above but for the periodicity fixed to be equal to $\pi$, the coefficients are given by:
\begin{equation}
\gamma_n=\frac{\Pi}{\pi}\int_{-\pi}^\pi \cos(2 n \eta) \sqrt{k^2+m^2 a^2(\eta)}.  
\end{equation}
Let us now consider the two limits: $k\gg m^2 a^2(\eta)$ and $k\ll m^2 a^2(\eta)$.
In the former case, one can expand the square root, obtaining the following formula for $\gamma_n$:

\begin{eqnarray}
\frac{\gamma_{n}}{\Pi}&\approx&\frac{1}{\pi}\int_{-\pi}^\pi \cos(2 n \eta) k\ d\eta \nonumber \\
&+&  \frac{m^2}{2 k} \sum_{i,j=0}^\infty \frac{1}{\pi}\int_{-\pi}^\pi b_i b_j \cos(2 n \eta) \cos(2 i \eta) \cos(2 j \eta)) d \eta \nonumber \\
&=&\delta_{n,0} \frac{k}{2} \nonumber \\
&+&\frac{m^2}{2\pi k}  \sum_{i,j=0}^\infty b_i b_j \int_{-\pi}^\pi \cos(2 n \eta) \cos(2 i \eta) \cos(2 j \eta)) d\eta \nonumber \\
&=&\delta_{n,0} \frac{k}{2} \nonumber \\
&+&\frac{m^2}{2\pi k} \sum_{i,j=0}^\infty b_i b_j
\frac{1}{4}[\delta_{n,i-j}+\delta_{n,i+j}+\delta_{n,j-i}+\delta_{n,-i-j}]
\nonumber~.
\end{eqnarray}    
Now the last delta contributes only if $n=i=j=0$, and thus:
\begin{eqnarray}
\frac{\gamma_{n}}{\Pi}&=&\delta_{n,0} (\frac{k}{2}+\frac{m^2}{8\pi k}b_0^2) \nonumber  \\
&+&\frac{m^2}{8\pi k} \sum_{i,j=0}^\infty b_i b_j [\delta_{n,i-j}+\delta_{n,i+j}+\delta_{n,j-i}] \nonumber \\
&=&\delta_{n,0} (\frac{k}{2}+\frac{m^2}{8\pi k}b_0^2) +\frac{m^2}{8\pi k} [ 2
\sum_{i=0}^\infty b_i b_{n+i}+\sum_{i=0}^{n-1}  b_i b_{n-i}]~, \nonumber \\
\end{eqnarray}    
so, if we define $\tilde k=\max_{\eta} m^2 a^2(\eta)$, we have:
\begin{eqnarray}
\frac{\gamma^{k\gg\tilde k}_{0}}{\Pi}&=&\omega^2=\frac{k}{2}+\frac{m^2}{8\pi k}\sum_{i=0}^\infty b_i^2  \nonumber  \\
\frac{ \gamma^{k\gg\tilde k}_{n\ge 1}}{\Pi} &=& \frac{m^2}{8\pi k} [ 2
\sum_{i=0}^\infty b_i b_{n+i}+\sum_{i=0}^{n-1}  b_i b_{n-i}]~.
\end{eqnarray}    

Due to the Parseval's identity, $\sum_{i=0}^\infty b_i^2=\frac{1}{2
\pi}\int_{-\pi}^{\pi} a^2(\eta) d\eta=\frac{1}{2 \pi}\int_{-\pi}^{\pi} a^2(\eta)
d\eta=E$. It is easy to see that for certain values of $\gamma_0$'s, $\alpha$ is
ill-defined. These are all the values of $k$ for which $\gamma_0$ takes an
integer value $n$; an easy calculation shows that these values are given by
$k_n=\Pi(n\pm \sqrt{n^2-\frac{m^2 E}{2\pi}})$. Since we are in the approximation in
which $k^2\gg m^2 a^2(\eta)$, we must have:
$k_{n\pm}=\Pi (n\pm n\sqrt{1-\frac{m^2 E}{2\pi n^2}})\approx n \Pi [1\pm(1-\frac{m^2
E}{4\pi n^2})]=\Pi*\{2n-\frac{m^2 E}{4\pi n^2},\frac{m^2 E}{4\pi n^2}\}$. Of these
phases, we disregard $k_{n-}$ as this violates the condition we started from.
Thus, these phases contain information on the whole evolution of the Universe's
scale factor $a^2(\eta)$, and are those values for which the Floquet's theory fails and the phase $\alpha$ becomes an integer number.

\end{document}